# Probing lactate exchange and compartmentation in Gray Matter via time-dependent diffusion-weighted MRS

Eloïse MOUGEL[1], Marco PALOMBO[2,3], Julien VALETTE[1]

[1] *Université Paris-Saclay, Commissariat à l'Energie Atomique et aux Energies Alternatives (CEA), Centre National de la Recherche Scientifique (CNRS), Molecular Imaging Research Center (MIRCen), Laboratoire des Maladies Neurodégénératives, 92260 Fontenay-aux-Roses, France*

[2] *Cardiff University Brain Research Imaging Centre (CUBRIC), School of Psychology, Cardiff University, Cardiff, UK*

[3] *School of Computer Science and Informatics, Cardiff University, Cardiff, UK*

**Corresponding author information:** Eloïse Mougel, Université Paris-Saclay, Commissariat à l'Energie Atomique et aux Energies Alternatives (CEA), Centre National de la Recherche Scientifique (CNRS), Molecular Imaging Research Center (MIRCen), Laboratoire des Maladies Neurodégénératives, 92260 Fontenay-aux-Roses, France. *E-mail:* eloise.mougel@cea.fr

**Present address corresponding author:** Commissariat à l'Energie Atomique et aux Energies Alternatives (CEA), Centre National de la Recherche Scientifique (CNRS), Molecular Imaging Research Center (MIRCen), Laboratoire des Maladies Neurodégénératives, 92260 Fontenay-aux-Roses, France.

**Abstract**

Lactate is crucial in the brain, as it is involved in neuronal activity and memory formation. This is thought to be related to the astrocyte-neuron lactate shuttle hypothesis (ANLS), which has been debated for 30 years, not least because it is difficult to measure lactate compartmentation and exchange *in vivo*. In particular, while ANLS requires transfer of lactate via the extracellular space, intracellular/extracellular exchange rate remains unknown. In this work, we propose to use time-dependent diffusion MRS to assess lactate exchange and lactate compartmentation in mouse brain gray matter *in vivo*. First, by comparing lactate time-dependent diffusivity and kurtosis with those of water and purely intracellular metabolites (which exhibit "fast" and "no" exchange signatures, respectively), we estimate that lactate exchange is slow (i.e. on the order of hundreds of milliseconds). Then, using biophysical models neglecting exchange, we estimate extracellular, neuronal and astrocytic lactate fractions compatible with the ANLS hypothesis.

# 1. Introduction

Lactate is considered a major component of neurometabolism in glutamatergic neurons through the astrocyte-neuron lactate shuttle (ANLS) mechanism (Pellerin & Magistretti, 1994). In particular, ANLS seems of fundamental importance in brain activity and memory formation. According to this hypothesis, glutamate release from neurons in the extracellular space activates glycolysis in astrocyte, leading to lactate production, which is then supplied to activated neurons. A concentration gradient from astrocytes to neurons across the extracellular space sustains this lactate shuttling mainly *via* monocarboxylate transporters (MCT). Disruption in the expression of certain transporters such as MCT2 in rat cortex, has been shown to impair memory formation (Roumes et al., 2021; Suzuki et al., 2011). However, ANLS remains controversial, as on the one hand the rate of lactate exchange remains unknown, and on the other hand it is difficult to measure lactate compartmentation non-invasively *in vivo* (Mächler et al., 2016). Measuring exchange rate and compartmentation non-invasively *in vivo* could be valuable information to underpin ANLS hypothesis.

Based on the detection of endogenous molecules, diffusion-weighted magnetic resonance spectroscopy (dMRS) is a powerful non-invasive tool that provides information on cell-specific microstructure (Palombo, Shemesh, et al., 2018; Ronen & Valette, 2015) and compartmentation of metabolite between intra- and extracellular spaces (Pfeuffer et al., 2000; Van Zijl et al., 1991). For the specific case of lactate, previous experiments confirm that dMRS allows distinguishing intra and extracellular spaces. Signal attenuation is specifically influenced by the extracellular contribution, as observed by Vincent et al. (Vincent et al., 2021), which may be used to separate intracellular from extracellular lactate when performing high *b*-value measurements (Pfeuffer et al., 2000), including in pathological conditions where lactate distribution between intracellular and extracellular spaces is altered (Malaquin et al., 2024). Furthermore, dMRS signatures of purely intracellular metabolites also exhibit some significant differences for metabolite diffusion in different cell type such as astrocytes and neurons (Palombo et al., 2016), thus it could also be possible to separate astrocytic and neuronal contribution in lactate signal.

Measuring dMRS as a function of diffusion time (like in time-dependent dMRI) may allow probing exchange (Aggarwal et al., 2020; Fieremans et al., 2010; Jelescu et al., 2022; Lee et al., 2020; Li et al., 2017). In gray matter, it has already been shown that kurtosis as a function of diffusion time for intracellular metabolites ($K_M(t_d)$) or water ($K_W(t_d)$) behaves differently, with an increase in kurtosis for intracellular metabolites due to the restricted environment in which metabolites diffuse, and a decrease in kurtosis for water which is in exchange between intra- and extracellular spaces (Mougel et al., 2024). The main difference between K($t_d$) behavior of intracellular metabolites and water arises from the presence or absence of exchange. The absence of exchange means that the diffusion is influenced by restriction within a compartment, i.e. within the cell for intracellular metabolites, resulting in diffusion that deviates from Gaussian diffusion with increasing $t_d$. As suggested by water diffusion simulation from other studies (Aggarwal et al., 2020; Lee et al., 2024; Novikov et al., 2023; Wu et al., 2024), at very short diffusion times, water exchange is constrained by cell membranes/barriers (due to limited exchange), resulting in a similar behavior to that of metabolite kurtosis, with an increase in kurtosis. In these simulations, again, when $t_d$ is long relative to the exchange time, exchange is no longer limited by cell membranes: diffusion should approach Gaussian diffusion (without being fully Gaussian) and a decrease in kurtosis is expected. The longer the exchange time, the longer the diffusion time must be to measure a decrease in kurtosis. Time-dependence of lactate kurtosis could therefore be used to estimate the rate of lactate exchange, or at least should make it possible to assess whether exchange significantly influences diffusion during $t_d$.

The aim of this paper is to assess lactate exchange and compartmentation in gray matter (GM) by measuring, for the first time, time-dependent diffusivity and kurtosis of lactate ($D_L(t_d)$/$K_L(t_d)$) *in vivo*. Firstly, the lactate exchange rate in GM is qualitatively assessed by comparison with time-dependent diffusivity and kurtosis of water ($D_W(t_d)$/$K_W(t_d)$) and intracellular metabolites ($D_M(t_d)$/$K_M(t_d)$) as measured in the same voxel, which exhibit signatures of "fast" (i.e. on the order of tens of ms (Jelescu et al., 2022; Mougel et al., 2024; Olesen et al., 2022; Uhl et al., 2024)) and "no" exchange respectively. Secondly, a simple biophysical model is built, with an assumption determined from the previously revealed information on lactate exchange and using parameters extrapolated from $D_M(t_d)$/$K_M(t_d)$ and $D_W(t_d)$/$K_W(t_d)$, to estimate extracellular, neuronal and astrocytic lactate fractions in GM of healthy mice.

## 2. Material and methods

### 2.1. dMRS acquisition

#### 2.1.1. Animal experimentation

All experimental protocols were reviewed and approved by the local ethics committee (CETEA N°44), and authorized by the French Ministry of Education and Research. They were performed in an approved facility (authorization #E92-032-02), in strict accordance with recommendations of the European Union (2010-63/EEC). All efforts were made to minimize animal discomfort. Animal care was supervised by veterinarians and animal technicians. Mice were housed under standard environmental conditions (12-hour light-dark cycle, temperature: 22 ± 1 °C and humidity: 50%) with *ad libitum* access to food and water.

### *2.1.2. Scanner, sequence and settings*

Experiments were performed on an 11.7 T BioSpec Bruker scanner interfaced to PV6.0.1 (Bruker, Ettlingen, Germany). A quadrature surface cryoprobe (Bruker, Ettlingen, Germany) was used for transmission and reception. Wild-type C57BL/6 mice were anesthetized with ~1.3% isoflurane and maintained on a stereotaxic bed with one bite and two ear bars. Throughout the experiment, body temperature was monitored and maintained at ~37 °C by warm water circulation. Breathing frequency was monitored using PC – SAM software (Small Animal Instruments, Inc., Stony Brook, NY).

The sequence used for all acquisitions consisted of a diffusion-weighted stimulated-echo block followed by LASER localization (STE-LASER) (Ligneul et al., 2017). Echo time was set to TE = 33.4 ms (including the 25-ms LASER echo time) and was held constant for each acquisition. *b*-values were interleaved between each repetition, and the gradient direction were changed to perform the so-called powder averaging over a number of directions equal to the number of repetitions. The duration of the diffusion gradient pulses was set to $\delta = 3$ ms. Diffusion gradients were separated by the delay $\Delta$, which is varied according to the protocol described in the next two sections (§2.1.3 and §2.1.4).

Water and metabolites diffusion was measured in six mice. A $7 \times 3 \times 3$ mm$^3$ (= 63 µL) spectroscopic voxel covering the two hemispheres was centered in the middle of the brain (incorporating midbrain, hippocampus and striatum) so that it contained mainly gray matter. Voxel composition (gray matter~90%, white matter~5% and CSF~5%) was estimated a posteriori on anatomical images acquired with a RARE sequence with TE/TR = 30/2500 ms, 78.1-µm isotropic resolution and 0.5-mm slice thickness, using manual segmentation with the Fiji distribution of the ImageJ software.

First and second-order shimming was performed using the Bruker MapShim method, based on a $B_0$ map. Next, a local iterative shimming was done in the voxel defined above. Linewidth is then $19.7 \pm 1.2$ Hz. Reference power was set in the center of the voxel in a 3 mm strip parallel to the surface coil.

*2.1.3. Water acquisition*

Water diffusion was measured in the spectroscopic volume of interest, for each delay $\Delta$ = 21.8, 31, 43.5, 101, 251, 501 ms i.e. resulting respectively in diffusion time $t_d$ = 20.8, 30, 42.5, 100, 250, 500 ms. Attenuation was measured for $b$ = 0.2, 0.6, 1.2, 2.0, 2.5 ms/µm², with 8 repetitions (so on 8 directions). Repetition time was set to TR = 5100 ms.

*2.1.4. Intracellular metabolites and lactate acquisition*

For each $t_d$, attenuations of N-acetylaspartate (NAA), choline compounds (tCho), creatine + phosphocreatine (tCr), myo-inositol (Ins), taurine (Tau) were measured for $b$ = 0.2, 0.9, 1.9, 3.2, 4.3, 6.9, 8.0, 15.0 ms/µm² and up to 8 ms/µm² for lactate (Lac) on 32 directions (= number of repetition). Two blocks of 32 repetitions/directions interleaved between each $t_d$ and $b$ value were acquired for $t_d$ = 42.5 and 100 ms and three blocks of 32 repetitions/directions were acquired to improve the SNR (which is approximately proportional to $N_{repetitions}^{1/2}$) for $t_d$ = 250 and 500 ms. Repetition time was set to TR = 2500 ms. In addition, a VAPOR water-suppression module supplemented by an additional 21-ms hermite inversion pulse inserted during the mixing time was used for metabolite acquisition.

*2.2 Data processing and analysis*

*2.2.1. Data processing*

Individual scans were frequency- and phased-corrected before averaging. For water acquisitions, the integral of water peak was computed on a 0.5-ppm window centered on the peak maximum at each $b$ and $t_d$. For metabolites, each spectrum was analyzed with LCModel (Provencher, 2001). Experimental macromolecular (MM) spectra, acquired for each $t_d$ using a double inversion recovery module ($TI_1$ = 2200 ms and $TI_2$ = 770 ms) at $b$ = 10 ms/µm², were included in LCModel basis-sets containing simulated metabolite spectra.

*2.2.2. Data Analysis*

For each $t_d$, the diffusion-weighted signal S as a function of $b$ (see §2.1.3 and §2.1.4 for the $b$ value range used for each kind of metabolites) was fitted to estimate the $t_d$-dependent apparent diffusivity of water or metabolites ($D_{W/M}(t_d)$) and kurtosis ($K_{W/M}(t_d)$), using the equation below (Jensen & Helpern, 2003), for each $t_d$:

$$S(b) = S(0) \times \exp(-bD + 1/6\, K\, b^2 D^2) \qquad \text{eq.[1]}.$$

### 2.2.2.1. Fitting $D_M(t_d)/K_M(t_d)$ to determine intracellular characteristic

As previously proposed in (Mougel et al., 2024), we used a very simple intracellular model to assess the influence of the restricted environment on intracellular metabolite diffusion. According to this paper, $D_M(t_d)/K_M(t_d)$ can be fitted by a model representing intracellular diffusion comprising two compartments, one representing soma by spheres and the other representing the projections by isotropically distributed cylinders, in a 20:80 ratio. Cylinders can also be replaced by sticks (meaning zeros-radius cylinder), as in (Malaquin et al., 2024; Palombo et al., 2020) (see supplemental information (**SI**) **Fig. S2** for the comparison of different models). First, the 20:80 ratio was imposed as in previous study (Malaquin et al., 2024; Mougel et al., 2024), then let free by simplifying the model with a stick model to avoid overfitting the data. An in-house iterative algorithm was used to fit models to D and K of each intracellular metabolite. This allowed extracting the mean intracellular diffusivity of metabolites, $D_{intra}$, and the mean radius of the soma, $R_{sphere}$, (and projections, $R_{cylinder}$, where appropriate).

### 2.2.2.2. Characterizing $D_W(t_d)/K_W(t_d)$ decrease to determine which structural disorder dominates

$D_W(t_d)/K_W(t_d)$ should be described by a power law, according to an increasing number of studies (Lee et al., 2020; Novikov et al., 2014). To assess which structural disorder influences $D_W(t_d)/K_W(t_d)$ in this large spectroscopic voxel with 5% CSF and 5% WM, we tested four different power laws, like those tested in (Mougel et al., 2024). We considered models (Lee et al., 2020; Novikov et al., 2014) given by

$$D(t_d) = D_\infty + C_D\, t_d^{-\theta} \qquad \text{eq.[2]},$$

$$K(t_d) = K_\infty + C_K\, t_d^{-\theta} \qquad \text{eq.[3]},$$

with the constants $C_D$, $C_K$, $D_\infty = \lim_{t_d \to \infty} D$ and $K_\infty = \lim_{t_d \to \infty} K$ let as free parameters to be determined by fitting the experimental data and setting the "universal" dynamical exponent $\theta = (p+d)/2$ to 0.5 (representing 1D structural disorder with a dimensionality d = 1 and a structural exponent p = 0, which determines the structural universality class) or 1 (representing either 2D/3D structural disorder, with either p = 0 and d = 2 corresponding for example to random discs or p = -1 or d = 3 corresponding for example to random rods), respectively. This analysis allowed identifying if 1D or 2D/3D disorder dominates. As an alternative analysis, the exponent theta can also be let as free parameter. Finally, as another alternative we also investigated whether the extracellular space mostly consists of randomly oriented fibers, by assessing if, at long $t_d$, $D_W(t_d)$ and $K_W(t_d)$ exhibit a logarithmic singularity of the form (Burcaw et al., 2015):

$$D(t_d) = C_D \log\left(\frac{t_d}{t_c}\right)/t_d + D_\infty \qquad \textbf{eq.[4]},$$

$$K(t_d) = C_K \log\left(\frac{t_d}{t_c}\right)/t_d + K_\infty \qquad \textbf{eq.[5]}.$$

## 3. Results & discussion

### *3.1 Assessing lactate exchange*

#### *3.1.1 Intracellular metabolite time-dependence reflects the absence of exchange*

In our large spectroscopic voxel, the high SNR (> 40) even for the highest *b* and longest $t_d$ allows reliable LCModel analysis for all intracellular metabolites (Cramér–Rao lower bound < 6%) (**Fig. 1**). As expected (Mougel et al., 2024), $D_M$ decreases and $K_M$ increases for all intracellular metabolites (**Fig. 2**). In the present study, improved reliability of the $K_M$ allows us to clearly identify that $K_M$ of predominantly intraglial metabolites (i.e. tCho and Ins) increases like $K_M$ of other metabolites and reaches almost the same plateau as the other metabolites. This was not so clear in previous work due to the lower reliability of glial metabolite data. It is also more apparent that $D_M$ of intraglial metabolites reaches a lower plateau than that of intraneuronal metabolites, despite similar free diffusion coefficients when measured in solutions (e.g. for NAA and Ins). This significant difference in $D_M(t_d)$ behavior between each cell type should enable us to identify parameters quite specific to glial or neuronal cells (see. **Table S1.1.** in **SI** for statistical analysis). This is also in line with previous data (Figure S4 in Appendix of (Palombo et al., 2016)).

Kurtosis behavior can be very well explained by the restricted environment in which metabolites diffuse in the absence of exchange with the extracellular space, as previously simulated in (Ianus et al., 2021). Without resorting to lengthy Monte Carlo simulations, a very simple model of spheres and uniformly distributed sticks may suffice to describe diffusion in the intracellular compartment and to estimate certain characteristics of the intracellular compartment, for this study. As shown in SI (**Fig. S2**), simplifying the model by replacing cylinders with sticks (thin cylinders of zero radius) and leaving the fraction of sticks free fits well to $D_M(t_d)$ and $K_M(t_d)$ for each metabolite. In particular, this model gives less uncertainty on the radius of somas (**Fig. 2**). $R_{sphere}$ and $D_{intra}$ are consistent with literature in healthy GM whatever the fitting strategy (Ligneul et al., 2019; Malaquin et al., 2024; Mougel et al., 2024; Palombo et al., 2016, 2020). However, $f_{stick}$ is slightly above the 80% initial fraction (~85%). $D_{intra}$ appears higher for intraneuronal metabolite than for intraglial metabolite. In particular $D_{intra}$ of NAA (intraneuronal) is higher than $D_{intra}$ of Ins (intra-astrocytic/-glial), whereas it has previously been measured that $D_{free}$ of NAA and Ins are identical (~0.75 µm²/ms at 37°C (Malaquin et al., 2024)), suggesting that $D_{intra}$ is cell-type specific. In previous simulations, it was shown that $D_{intra}$ as estimated when modeling with cylinders can be influenced by features not accounted for in the model and that may be different from one cell type to another, e.g. $D_{intra}$ as extracted when modeling diffusion in infinite cylinders decreases when increasing spines/leaflets density (Palombo, Ligneul, et al., 2018). Such difference in the density of short-scale secondary structures (spines for neurons, leaflets for astrocytes) may be a plausible hypothesis to explain why $D_{intra}$ are different from one cell type to another.

### *3.1.2. Water time-dependence as reference for fast exchange between intra and extracellular spaces*

As expected, D and K decrease for water (**Fig. 4A**). Adding 5% WM and 5% CSF in the voxel composition does not radically alter the overall behavior of $D_W$ and $K_W$ as a function of diffusion time. The choice of a different voxel compared to previous work should therefore not affect previous conclusions. To verify this statement and estimate which class of disorder governs diffusion time dependence of $D_W$ and $K_W$ in this larger voxel, the four different power laws given by eq.[2]-[5] are fitted to $D_W$ and $K_W$, in line with previous work (Mougel et al.,2024). Law in -0.5 fits slightly better to the data (cAIC is lower, **Fig. 3**). Dimensionless ratio $\xi$ is rather similar (~1.5). Hence, 1D short-range disorder seems to dominate water diffusion time dependence. Conclusions are thus still valid in this large voxel.

Based on previous results (Mougel et al., 2024), water exchange time is presumably on the order of tens of ms (NEXI/SMEX cannot be applied on these data due to the range of *b* chosen in our study). These data confirm, as have many previous works, that diffusion is governed by non-Gaussian effects when diffusion time is long (Burcaw et al., 2015; Jelescu et al., 2022; Lee et al., 2020; Mougel et al., 2024; Novikov et al., 2014). The Kärger model is also not valid due to non-Gaussian diffusion (see **Fig. S3** in **SI**). It remains unclear if the power-law decay is mainly influenced by the diffusion in a non-Gaussian compartment or by exchange and what is an accurate estimates of the water exchange time.

### *3.1.3. Specific lactate time-dependence indicates slow exchange between ICS and ECS*

In our large voxel, lactate signal is high enough (SNR ~10) to allow a good fit using LCModel, yielding CRLBs < 8%, even for the highest $b$ and longest $t_d$ (**Fig. 1B**). Thanks to the specific range of $b$ up to 8 ms/µm², well adapted to fit the kurtosis representation to the data, standard deviations of the mean of $D_L$ and $K_L$ are low (**Fig. 4C**). As expected, ECS contribution is clearly visible on the estimated lactate parameters, with $D_L$ much higher than $D_M$ and $K_L$ lower than $K_M$ (**Fig. 4B-4C**).

Strikingly, $K_L(t_d)$ increases slightly, a far cry from the kurtosis behavior of water, the endogenous reference for measurable exchange (**Fig. 4**). The absence of decrease in $K_L(t_d)$ indicates that diffusion in a restricted environment still has some impact on the signal measured in this $t_d$ range. Therefore, compared to time dependencies of water and metabolites, the specific behavior for lactate suggests that the typical lactate exchange time between ICS and ECS is much longer than that of water, otherwise we should instead observe a decrease in kurtosis. According to water diffusion simulation (Figure 2-3 in (Novikov et al., 2023)), K should start decreasing for $t_d$ longer than the exchange time meaning that lactate exchange time must be longer than the probed $t_d$, since $K_L$ does not decrease. Under these conditions, exchange contribution should be negligible in lactate diffusion-weighted signal, at least over the probed $t_d$ in our experiments, suggesting that modeling lactate diffusion does not require accounting for exchange.

### *3.2. Assessing lactate compartmentation*

### *3.2.1 Three-compartment model*

Using a biophysical model to fit to $D_L(t_d)/K_L(t_d)$ may provide an estimate of lactate compartmentation in GM. In this study, $D_M(t_d)/K_M(t_d)$ as well as extracted parameters $D_{intra}$ and $R_{sphere}$ estimated for neurons and astrocytes show significant differences (**§3.1.1.**), so it seems possible to build a model incorporating two different intracellular contributions (neurons and astrocytes) as well as the extracellular contribution. Moreover, lactate exchange between compartments can be neglected at the diffusion times considered here (**§3.1.3.**). Inspired by the two-compartment model used in (Malaquin et al., 2024) to model lactate signal attenuation in high-*b* experiments, a new three-compartment model can be built by separating the intracellular contribution into two. Hence, three-compartment model used to model $D_L(t_d)/K_L(t_d)$ is given by fitting the eq.[1] to the signal model:

$$S(b, t_d) = f_{intraN} S_N(b, t_d) + f_{intraA} S_A(b, t_d) + f_{extra} S_{ECS}(b, t_d) \quad \text{eq.[7]}$$

where:
- intraneuronal contribution to the signal ($S_N$) is represented using spheres-and-sticks model, with parameters extracted from NAA, which is mainly present in neurons (i.e. $R_{sphereN}$ = 5.6 µm, $D_{intraN}$ of lactate = 0.47 µm²/ms in a ratio sphere:stick = 15:85). $D_{intraN}$ of lactate is determined from the $D_{intra}$ of NAA by considering that $D_{intra}$ of lactate is ~30% higher, due to the smaller size of lactate and based on the free diffusions of NAA and Lac measured in vitro by (Malaquin et al., 2024).
- intraastrocytic contribution to the signal ($S_A$) is represented using spheres-and-sticks model, with parameters extracted from Ins, which is mostly present in astrocytes (i.e. $R_{sphereA}$ = 7.60 µm , $D_{intraA}$ of lactate = 0.34 µm²/ms in a ratio sphere:stick = 15:85). For the same reason as stated previously, $D_{intraA}$ of lactate is also considered to be ~30% higher than $D_{intra}$ of Ins.
- extracellular compartment signal ($S_{ECS}$) is initially based on a Gaussian diffusion in the extracellular space, e.g. as used in (Malaquin et al., 2024), with $D_{extra}$ let as a free parameter.
- extracellular fraction ($f_{extra}$) and intraneuronal fraction ($f_{intraN}$) are let free, and intraastrocytic fraction ($f_{intraA}$) is defined as $f_{intraA} + f_{intraN} + f_{extra} = 1$.

### 3.2.2 Modeling lactate time-dependence allows separating neuronal and astrocytic lactate fractions

This three-compartment model fits well to $D_L/K_L$ for $t_d$ up to 250 ms. Estimated extracellular fraction is ~40%, which is compatible with the extracellular fraction estimated in previous works using a two-compartment model on data acquired at $t_d$ ~50 ms and up to high *b* value (~20 ms/µm²) in wild-type mouse cortex (Malaquin et al., 2024). Diffusivity in extracellular space is 0.52 ± 0.04 µm²/ms, which is also close to $D_{extra}$ estimated in Malaquin et *al*. At $t_d$ =

500 ms, D is overestimated by this model, but K is well fitted. Sensitivity analysis (**SI Fig. S5.2**) shows that the model is more sensitive to extracellular parameters ($D_{extra}$ and $f_{extra}$) than intracellular ones. ECS contribution is thus a crucial parameter in this model, and its impact is greater at longer $t_d$. Poor modeling of the ECS contribution may thus explain the overestimation of D at long $t_d$. In addition, this model fits slightly better to $D_L$ and $K_L$ than ICS-ECS two-compartment model used in Malaquin et al. as shown in **SI fig.S4.1**, where the ICS-ECS two-compartment is applied to our data (BIC is lower and cAIC is lower than or equal with the three-compartment model). Not surprisingly, as the definition of ECS contribution is the same in both models, D is overestimated with both models for the longest $t_d$. However, by splitting ICS contribution between astrocytic and neuronal contributions, a slight improvement is observed for the $K_L$ fit at the shorter $t_d$, and this model allows estimation of the fraction of intraneuronal lactate that contributes to the signal. Neuronal contribution represents ~4% ($\pm$ 4%, CI 95%) of the total lactate signal, while astrocytic contribution ~56% ($\pm$ 4.4%, CI 95%) and extracellular contribution ~40% ($\pm$ 0.3%, CI 95%). This is therefore compatible with a lactate gradient from astrocytes to neurons via extracellular space, in line with the ANLS hypothesis (Pellerin & Magistretti, 1994).

## 4. Strengths and Limitations

*Exchange and lactate modeling*

In this paper, exchange is only explored qualitatively, as new models would be required to estimate lactate exchange time. Models used in this work initially aim at illustrating the general behavior of $D_L$ and $K_L$, and at interpreting these results. As no decrease or inflection point is observed, there may be no way to estimate exchange time in this $t_d$ range, at least with current diffusion models in GM (Jelescu et al., 2022; Olesen et al., 2022). Next challenge may be to find a way of modeling lactate diffusion in GM that accounts for exchange, in order to estimate lactate exchange time that might be necessary for measurement at longer $t_d$.

Having said this, as the three-compartment model neglecting exchange fits well to the data, we can discuss a bit further the fundamental contributions made by this study. Firstly, these observations are crucial for lactate modeling, as they suggest that lactate exchange time is negligible at least for the shortest diffusion time tested in this experiment ($t_d$<250 ms). For this reason, we have tested in **SI fig.S4.1** an ICS-ECS two-compartment model neglecting exchange and in the main text a slightly more refined three-compartment model neglecting exchange, both of which fit well to our data. The ICS-ECS two-compartment model neglecting exchange can be used to model lactate diffusion as a first approximation for this range of $t_d$, although the three-compartment model gives better results. In addition, this study shows that a SANDI-type model (Palombo et al., 2020) comprising sticks and cylinders, adapted to intracellular metabolites, can be easily transposed to the case of lactate. Lactate diffusion in extracellular space, on the other hand, is much more complex to model. Non-Gaussian diffusion in extracellular space may be necessary to refine the model, as we can see in the sensitivity analysis, which suggests that the extracellular contribution strongly influences both models (**SI fig S4.3, S5.2**). However, characterizing the diffusion properties in ECS remains a challenge. This could for example be achieved by using an exogenous probe (Vincent et al., 2021). Finally, regarding results of the three-compartment model, as the lactate fraction in neurons appears to be very low, we could also simplify the model by considering only astrocytic and extracellular compartments (**Fig. 8**). Unsurprisingly, this model gives a smaller BIC, as the number of free parameters in this new astrocyte-ECS two-compartment model is smaller, cAIC is slightly lower too. This model yields the same extracellular fraction, but is less useful in the context of ANLS.

*In the context of ANLS*

Separating ICS in a neuronal compartment and an astrocytic compartment to model lactate diffusion might be highly relevant, especially in the context of ANLS. However, this model may not be systematically applied to all lactate diffusion data and may not give reliable information if $D_L(t_d)/K_L(t_d)$ are too noisy (noise > 25%), as shown in the model sensitivity analysis in **SI S5.1**. First, this three-compartment model is only applicable if the small differences between diffusion in astrocytes and neurons are detectable. Here we measured a clear difference between D of Ins and NAA and this difference is even more marked at longer $t_d$ (~10% lower for Ins at shorter $t_d$ and ~25% at longer $t_d$). Acquiring data at multiple $t_d$ may therefore facilitate separation between astrocytic and neuronal diffusion properties. Moreover, it is interesting to note that the estimated intracellular fractions between astrocytes and neurons are in line with ANLS: indeed a larger fraction of lactate signal comes from astrocytic compartment (~56% of the total lactate signal), and a minor contribution from the neuronal compartment (~4% of the total lactate signal). In the context of ANLS, a valuable information would be the concentration gradient. Considering volume fractions of each compartment (i.e. 45% of neurons, 25% of astrocytes, 20% of ECS in GM (Aubert & Costalat, 2005; Hertz, 2008)) and assuming that total lactate signal corresponds to a concentration of ~1 mM in GM, we can estimate that $[Lac]_N$ ~0.09 mM ($\pm$ 0.9 mM, CI 95%), $[Lac]_A$ ~2.2 mM ($\pm$ 0.14 mM, CI 95%) and $[Lac]_{ECS}$ ~ 2 mM ($\pm$ 0.02 mM, CI 95%), which is in line with the literature (Mosienko et al., 2015). This is also in agreement with a gradient of concentration from astrocytes to neurons across the ECS as required for ANLS, and as measured previously during invasive FRET experiments (Mächler et al., 2016). This study is a first step in the non-invasive observation of lactate concentration gradient from astrocytes to neurons, opening up new perspectives for the application of time-dependent DW-MRS to determine compartmentation between neurons, astrocytes and ECS.

This work focuses on lactate diffusion in the healthy brain, but things may be different in pathological conditions. For example, in the APP/PS1 mouse model of Alzheimer disease, decreased extracellular lactate fraction was measured with high *b*-value measurements at short $t_d$ (~50 ms) and confirmed with implanted enzyme-electrodes (Malaquin et al., 2024), which was proposed to be associated with decreased ANLS. However, neuronal *versus* astrocytic lactate fractions were not assessed in that previous work, so the potential decrease of astrocytic fraction, likely associated with decreased extracellular fraction and decreased ANLS, was not confirmed. It would be informative, in such context, to perform time-dependent DW-MRS experiments to extract information about lactate compartmentation between neurons, astrocytes and ECS. Furthermore, compared with other methods such as high *b*-value measurement, the *b*-value range required to fall within the validity range of the kurtosis representation is compatible with the usual ~50 mT/m gradient of most clinical scanners. Such time-dependent measurements are time-consuming, but limiting the measurement to two suitably chosen $t_d$ might make this method applicable to humans.

## 5. Conclusion

To our knowledge, this work is the first reporting the time-dependent $D_L/K_L$ measurements. These measurements show that $K_L$ increases slightly, meaning that cell "barriers" still have some impact on diffusion, thus providing experimental evidence that lactate exchange is slow compared to the diffusion times probed, and likely on the order of hundreds of milliseconds. Consequently, to model lactate diffusion in GM, at least on the diffusion time range probed in this experiment (40-500 ms), it is not necessary to take exchange into account, and a simple two-compartment model may be sufficient as a first approximation. An alternative model considering neuronal, astrocytic and extracellular compartments is also applicable to these time-dependent measurements, and allows for the first time to estimate astrocytic and neuronal fractions compatible with ANLS hypothesis. Time-dependent DW-MRS appears to be a powerful tool for quantifying extracellular fraction and disentangling contributions of astrocytes and neurons to the intracellular lactate signal. This method holds promise for further application in pathological conditions.

**Data and Code Availability:**

Data will be deposited before the publication in Zenodo.

**Author contributions:**

Conceptualization: EM, MP, JV

Data curation: EM

Formal Analysis: EM

Methodology: EM

Funding acquisition: MP, JV

Writing original draft: EM, JV

Writing review and editing: EM, MP, JV

**Funding:** This project has received funding from the European Research Council (ERC) under the European Union's Horizon 2020 research and innovation programmes [grant agreement No 818266] and the UKRI Future Leaders Fellowship MR/T020296/2. The 11.7 T MRI scanner was funded by a grant from "Investissements d'Avenir - ANR-11-INBS-0011 - NeurATRIS: A Translational Research Infrastructure for Biotherapies in Neurosciences". M.P. is supported by the UKRI Future Leaders Fellowship MR/T020296/2.

**Declaration of competing interests:** The authors declare no competing financial interests.

**Supplementary Material:** see SI_Kurtosis_Lactate_V2.docx

**Figures**

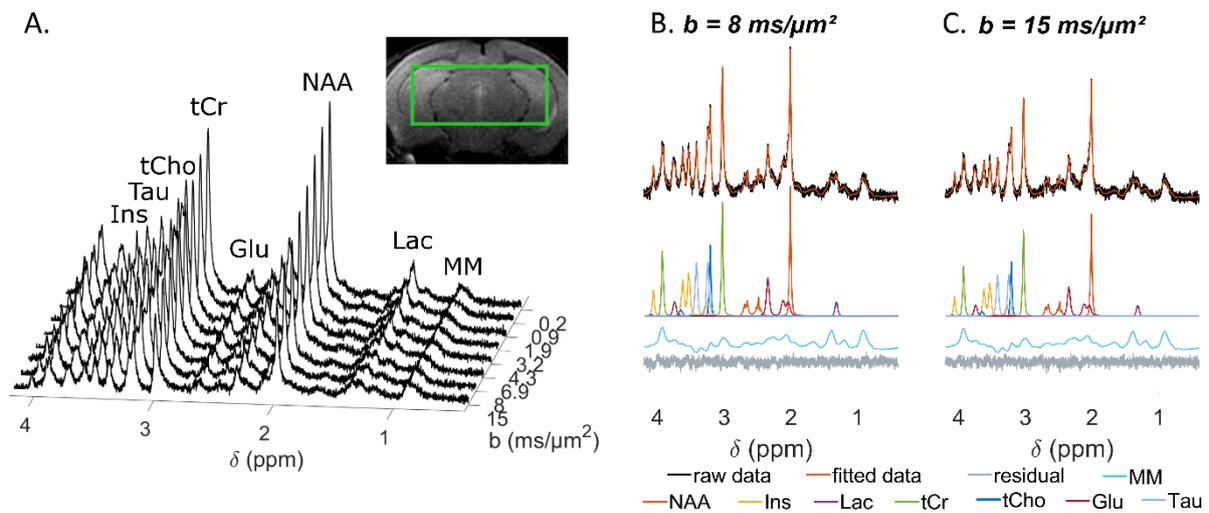

**Fig. 1.** Example of signal attenuation at $t_d$ = 500 ms and LCModel decomposition of spectra acquired in the 63-µL voxel in one mouse brain. A. Stack-plot showing typical signal attenuation as a function of $b$ value for that $t_d$. Image shows the position of the spectroscopic voxel, which mainly contains gray matter (90% GM, 5% WT, 5% CSF). B-C. Graphs display the LCModel decomposition at $b$ = 8 ms/µm² and 15 ms/µm², which are the highest $b$ values (i.e. lowest SNR spectra) for the fit of the signal attenuation of lactate and intracellular metabolites, respectively.

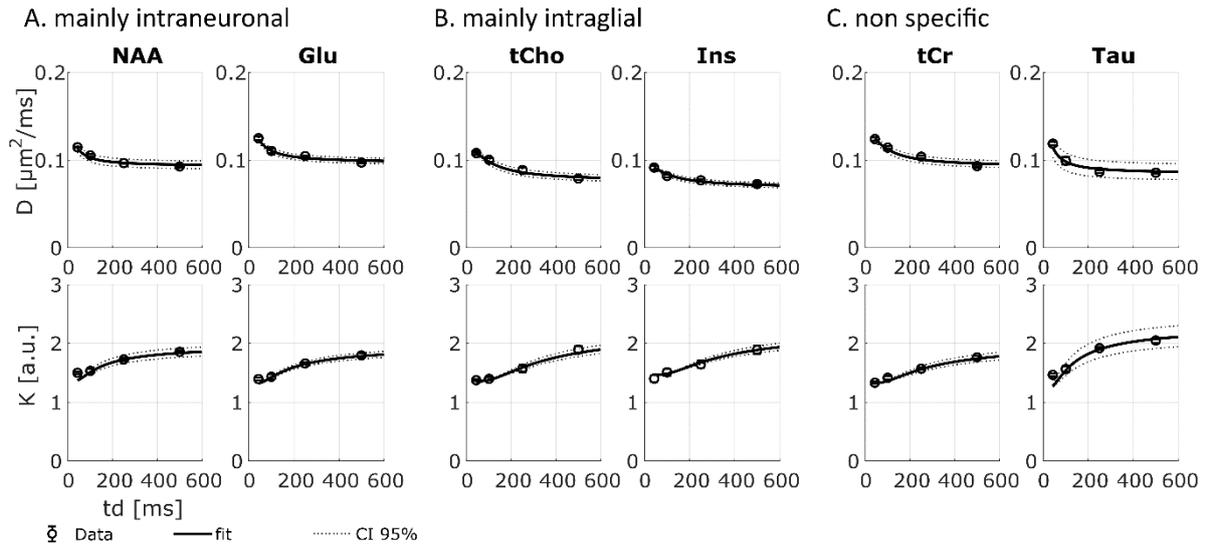

**Fig. 2.** $D_M$ and $K_M$ as a function of $t_d$ fitted with a biophysical model for each intracellular metabolite. A-C. Metabolites data are grouped according to whether they are found mainly in neurons or glial cells or not specifically in a given cell type. D. Biophysical model used to fit to the data models the intracellular compartment by spheres and sticks representing somas and projections, with intracellular diffusivity ($D_{intra}$), sphere radius ($R_{sphere}$) and stick fraction ($f_{stick}$) being let as free parameters.

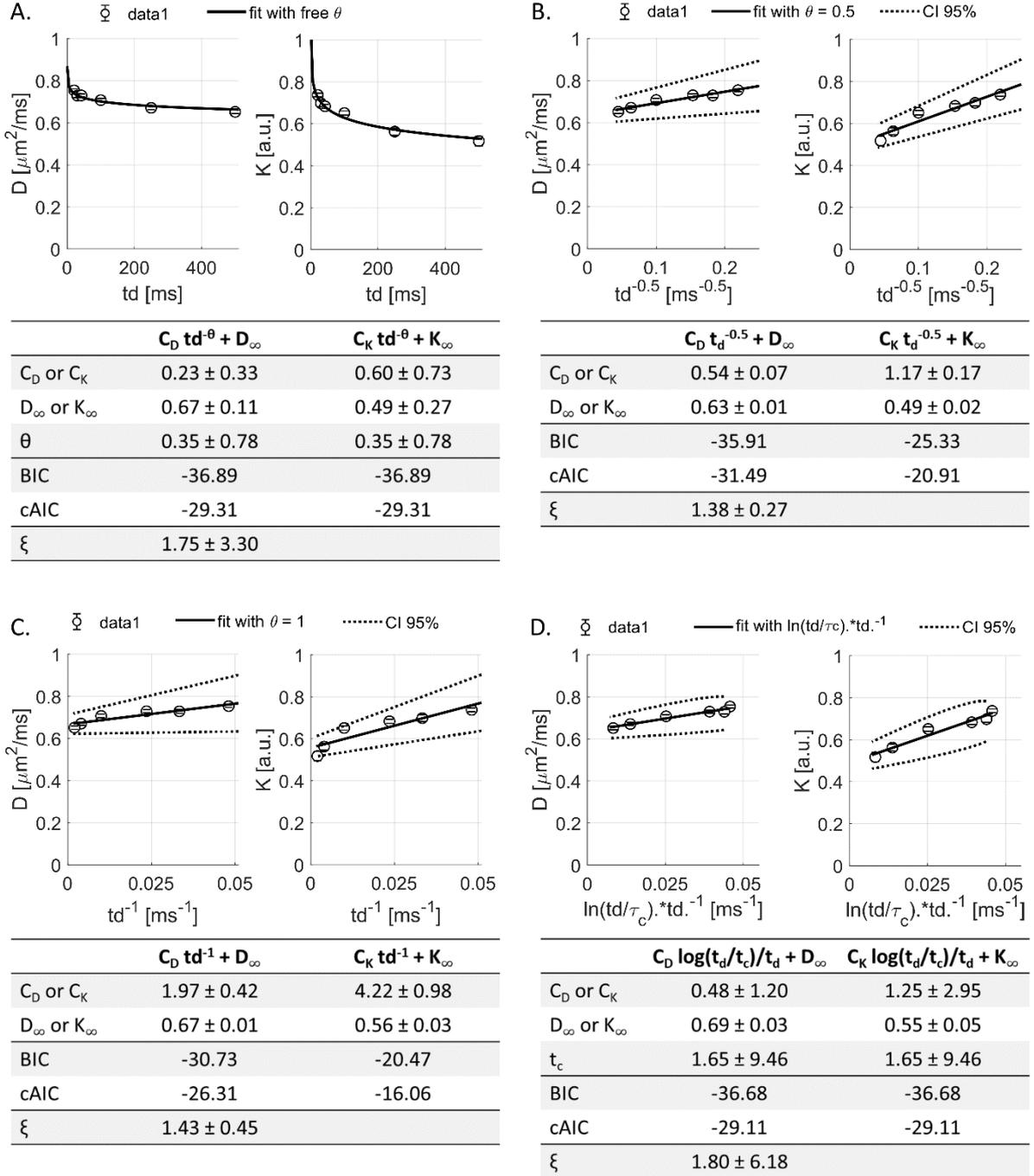

**Fig. 3.** D and K of water as a function of $t_d$ fitted with different decay laws. A. Decay law given by eq.[2] and eq.[3] fits well to $D_W(t_d)$ and $K_W(t_d)$. $D_W(t_d)$ and $K_W(t_d)$ are jointly fitted, and exponent $\theta$ is let as free parameter. B. Decay law given by eq.[2] and eq.[3], with a fixed $\theta=0.5$ exponent, corresponding to 1D-short range structural disorder, fits well to $D_W(t_d)$ and $K_W(t_d)$, respectively. C. Decay law given by eq.[2] and eq.[3], with a fixed $\theta=1$ exponent, corresponding to 2D/3D-short range structural disorder, fits less well to $D_W(t_d)$ and $K_W(t_d)$ respectively. D. Decay law given by eq.[4] and eq.[5], corresponding to some special 2D-short range disorder, fits less well to $D_W(t_d)$ and $K_W(t_d)$ respectively.

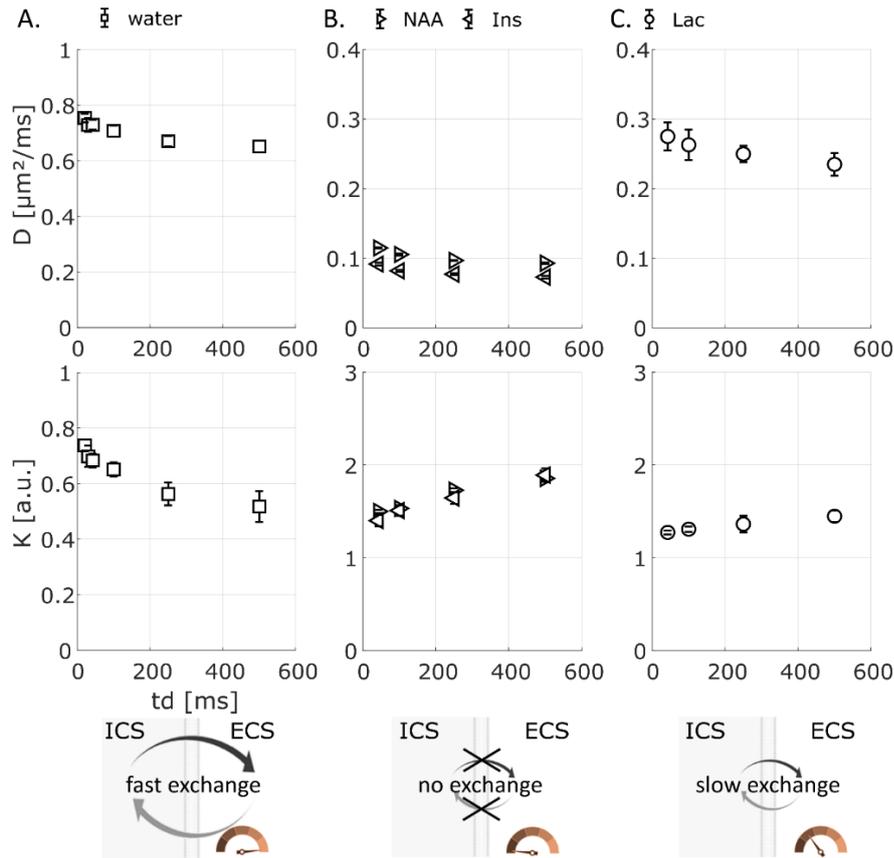

**Fig. 4** Mean and standard error of the mean over six mice of D and K as a function of $t_d$ for water (A), two intracellular metabolites (B) and lactate (C). $K_W$ (squares) decreases due to fast exchange, while $K_M$ (triangles) and $K_L$ (circles) increase. $K_M$ clearly rises, consistently with intracellular metabolites being confined in a restricted environment (no exchange). $K_L$ slightly increases, which means that restriction inside the cells still has an effect and that exchange time is rather long compared to water (slow exchange), as no decrease is observed on the diffusion time range.

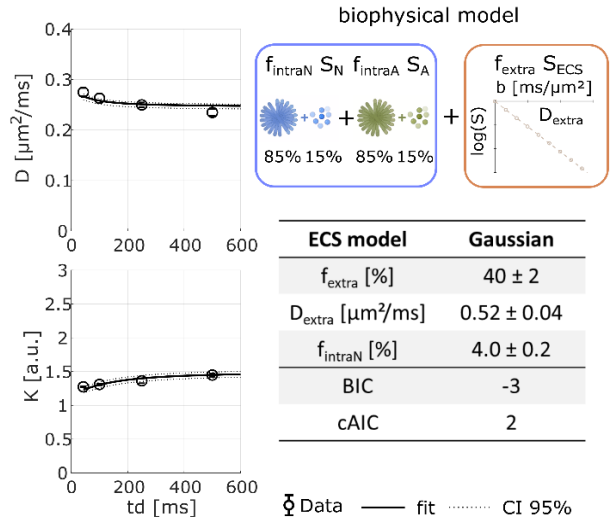

**Fig. 5.** Three-compartment model fits well to $D_L(t_d)/K_L(t_d)$, but there is an overestimation at long $t_d$. The model includes three contributions: neuronal and astrocytic contributions (decomposed into spheres and sticks) with parameters estimated for NAA and Ins diffusion, respectively (i.e. $R_{sphereN}$ = 5.6 µm, $D_{intraN}$ = 0.47 µm²/ms; $R_{sphereA}$ = 7.6 µm, $D_{intraA}$ = 0.34 µm²/ms, and $f_{stick}$ = 85%), and ECS compartment represented by Gaussian diffusion. $f_{extra}$ is a free parameter.

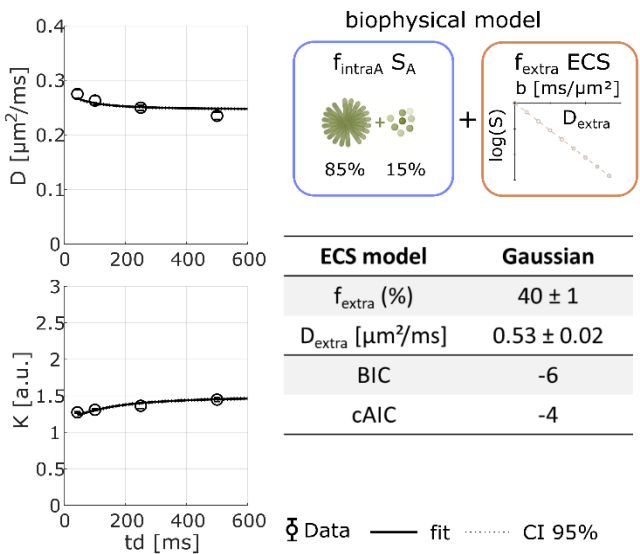

**Fig. 6.** Astrocyte-ECS two-compartment model fits well to $D_L(t_d)/K_L(t_d)$, while there is still an overestimation at long $t_d$. The two compartments consist of an astrocytic compartment composed of spheres and cylinders estimated from Ins diffusion (i.e. $R_{sphereA}$ = 7.6 µm, $D_{intraA}$ = 0.34 µm²/ms, and $f_{stick}$ = 85%), and ECS compartment represented by Gaussian diffusion. $f_{extra}$ is a free parameter.